\documentclass[12pt]{article}
\usepackage{amsmath}
\usepackage{graphicx}
\usepackage{subcaption}

\newcommand\pubnumber{DPF2013-112}
\newcommand\pubdate{\today}

\newcommand\nova{NO$\nu$A }
\newcommand\nue{$\nu_e$ }
\newcommand\numu{$\nu_{\mu}$ }

\graphicspath{{./}}
\def\napoli{School of Physics and Astronomy\\
University of Minnesota, Twin Cities, MN-55455, USA}
\def\support{\footnote{Work supported by Department Of Energy grant DE-FC02-07ER41471}}

\def\Title#1{\begin{center} {\Large #1 } \end{center}}
\def\Author#1{\begin{center}{ \sc #1} \end{center}}
\def\Address#1{\begin{center}{ \it #1} \end{center}}

\newcommand\pubblock{\rightline{\begin{tabular}{l} \pubnumber\\
         \pubdate  \end{tabular}}}
\newenvironment{Abstract}{\begin{quotation}  }{\end{quotation}}
\newenvironment{Presented}{\begin{quotation} \begin{center} 
             PRESENTED AT\end{center}\bigskip 
      \begin{center}\begin{large}}{\end{large}\end{center} \end{quotation}}

\def\beq{\begin{equation}}
\def\eeq#1{\label{#1}\end{equation}}
\def\eeqn{\end{equation}}

\def\beqa{\begin{eqnarray}}
\def\eeqa#1{\label{#1}\end{eqnarray}}
\def\eeqan{\end{eqnarray}}

\let\bar=\overbar

\def\Dslash{\not{\hbox{\kern-4pt $D$}}}
\def\dslash{\not{\hbox{\kern-2pt $\del$}}}

\def\msb{{\bar{\ssstyle M \kern -1pt S}}}

\begin{document}
\begin{titlepage}
\pubblock

\vfill
\Title{A Data-Driven Method of Background Prediction at \nova}
\vfill
\Author{ Kanika Sachdev\support}
\Address{\napoli}
\vfill
\begin{Abstract}
\nova is a long-baseline neutrino oscillation experiment that will use the NuMI beam originating at Fermilab. \nova enables the study of two oscillation channels: $\nu_{\mu}$ disappearance and $\nu_{e}$ appearance. It consists of two functionally identical detectors, the near detector (ND) at Fermilab and the far detector (FD) near International Falls in Northern Minnesota. The ND will be used to study the neutrino beam spectrum and composition before oscillation, and measure background rate to the $\nu_e$ appearance search. In this paper, I describe a data-driven technique to estimate the neutral current (NC) component of the ND spectrum. Using the $\nu_{\mu}$ CC interactions where the reconstructed muon is removed from the event, we produce a well understood sample of hadronic showers that resemble NC interactions.
\end{Abstract}
\vfill
\begin{Presented}
DPF 2013\\
The Meeting of the American Physical Society\\
Division of Particles and Fields\\
Santa Cruz, California, August 13--17, 2013\\
\end{Presented}
\vfill
\end{titlepage}
\def\thefootnote{\fnsymbol{footnote}}
\setcounter{footnote}{0}

\section{The \nova Experiment}
The NuMI Off-axis \nue Appearance (\nova ) experiment is designed to observe the appearance of electron neutrinos in the NuMI (Neutrinos at Main Injector) beam, which is primarily a muon neutrino beam from Fermilab. The \nova far detector is a 14kt liquid scintillator detector located near International Falls in Northern Minnesota, at a distance of 810km from the proton target at Fermilab. The near detector is 0.33kt, and is located close to the source at Fermilab. The detectors are constructed with liquid scintillator filled PVC modules divided into long tube-like cells. More details on \nova detector design can be found in \cite{Ayres:2007tu}.
\begin{figure}[htb]
  \centering
  \includegraphics[height=2in]{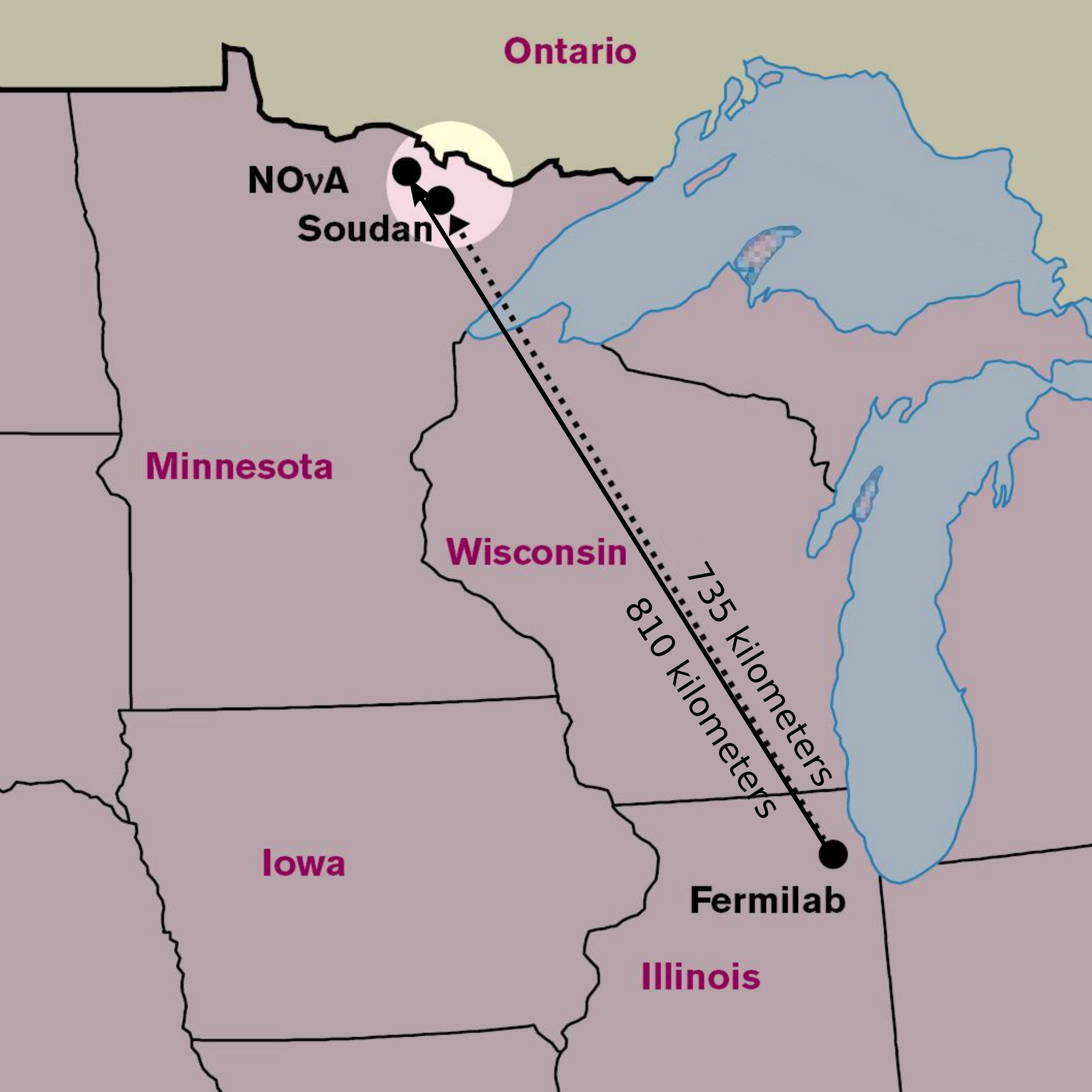}\qquad
  \includegraphics[height=2in]{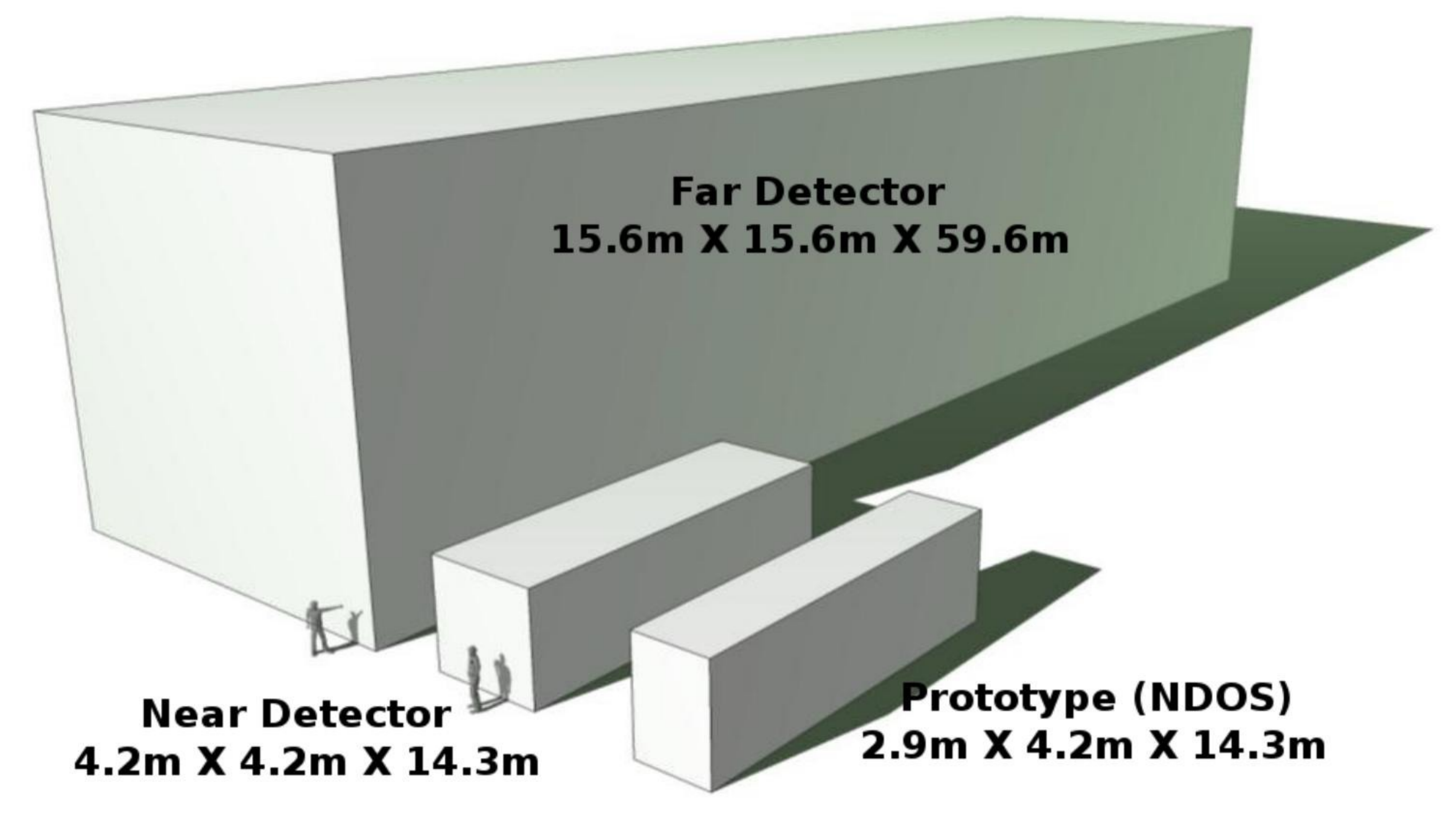}
  \caption{Graphics that show the placement and relative sizes of the \nova detectors.}
  \label{fig:nova}
\end{figure}

The off-axis placement of the \nova detectors provides a narrow-band beam peaked at 2 GeV, which, for this base-line, is close to the \numu $\rightarrow $ \nue oscillation maximum. The long baseline of the experiment ensures that interaction of the beam neutrinos in the earth will produce a significant matter effect. This unique placement of the \nova experiment gives us the ability to make precision measurement of oscillation parameters like $\theta_{23}$, $\Delta m_{32}^2$ and $\theta_{13}$ and probe the octant of $\theta_{23}$, the CP violation phase, $\delta$ and the neutrino mass hierarchy (is $m_3 > m_1,m_2$).

\section{Oscillation Analysis At \nova}
\begin{figure}[htb]
  \centering
  \includegraphics[width=\textwidth]{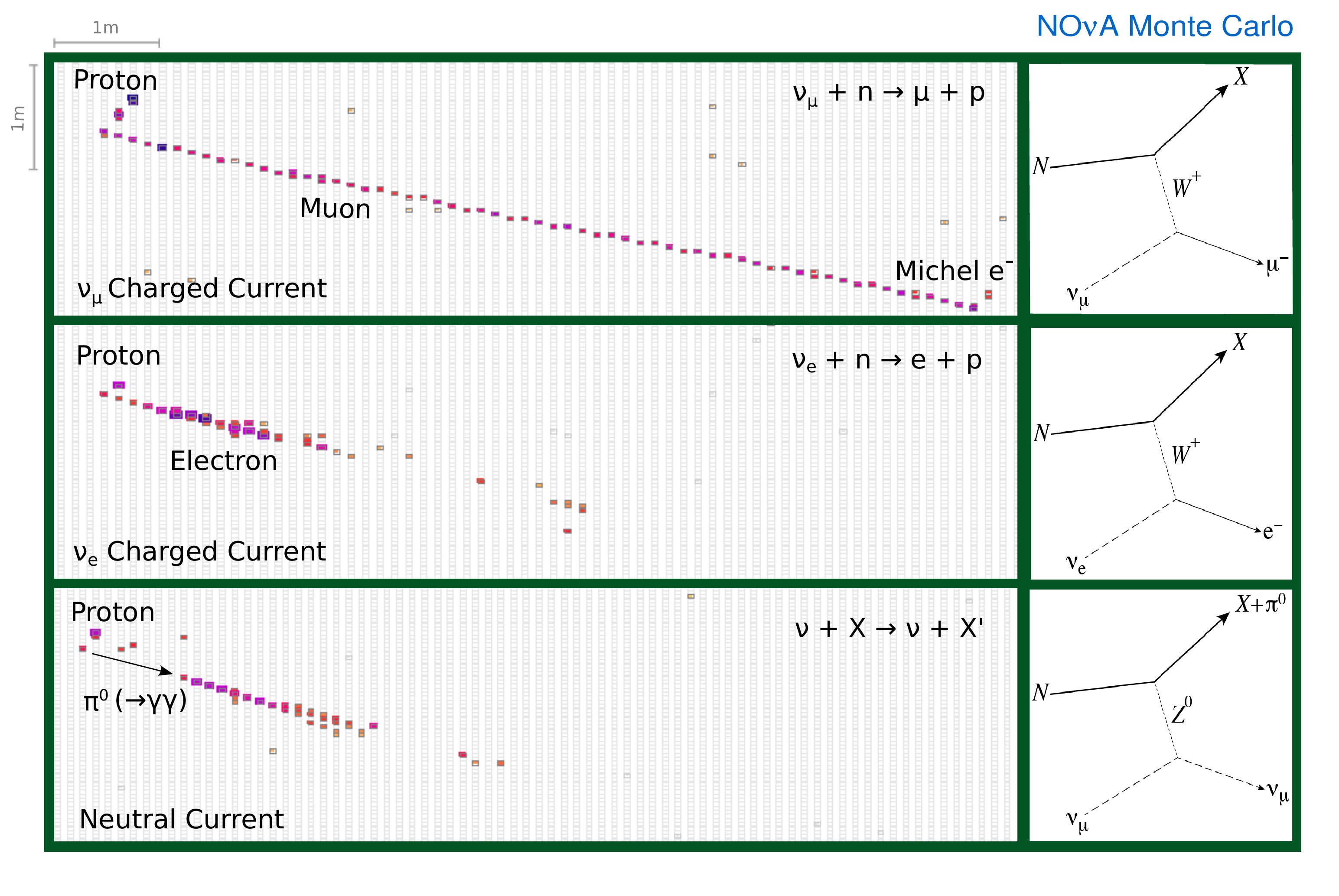}
  \caption{Event topologies in the \nova detectors\label{fig:evd}}
\end{figure}

The final state in a neutrino interaction consists of a lepton, a charged lepton in case of Charged Current (CC) or a neutrino in case of Neutral Current (NC), and a hadronic shower resulting from the recoil of the scattering nucleus. The \nova detectors have been designed to observe electromagnetic showers resulting from the electron in the final state of a charged current (CC) interaction of \nue. Muons in \nova appear as long clean tracks (see figure \ref{fig:evd}); therefore, \numu CC interactions are not a significant background to the \nue appearance search. The NC interactions form the major source of background because hadronic showers can occasionally be misidentified as electron showers.

The near detector, due to its proximity to the NuMI beam-line, will observe the beam when the neutrinos have not yet oscillated. It, therefore, offers a background-only data sample for the \nue analysis. The ND data will be decomposed into the constituent interactions: \numu CC, NC and the CC interactions of the small ($\sim2\%$) \nue component of the beam. Each of these interactions differ in how they propagate to the far detector: CC interactions will exhibit oscillations, NC's are flavor independent and so are unaffected by oscillations. The oscillation-corrected extrapolation of the decomposed ND spectra to the FD results in an estimate of the expected background rates in the \numu$\rightarrow$\nue oscillation analysis.

Thus, the near detector plays a central role in the oscillation analysis. Various methods are being implemented at \nova to estimate the contribution of the different interaction types to the observed spectrum in the ND. The rest of the paper describes one such technique called the Muon Removed Charged Current.

\section{Muon Removed Charged Current}
\begin{figure}
  \centering
  \begin{subfigure}[b]{0.27\textwidth}
    \includegraphics[width=\textwidth]{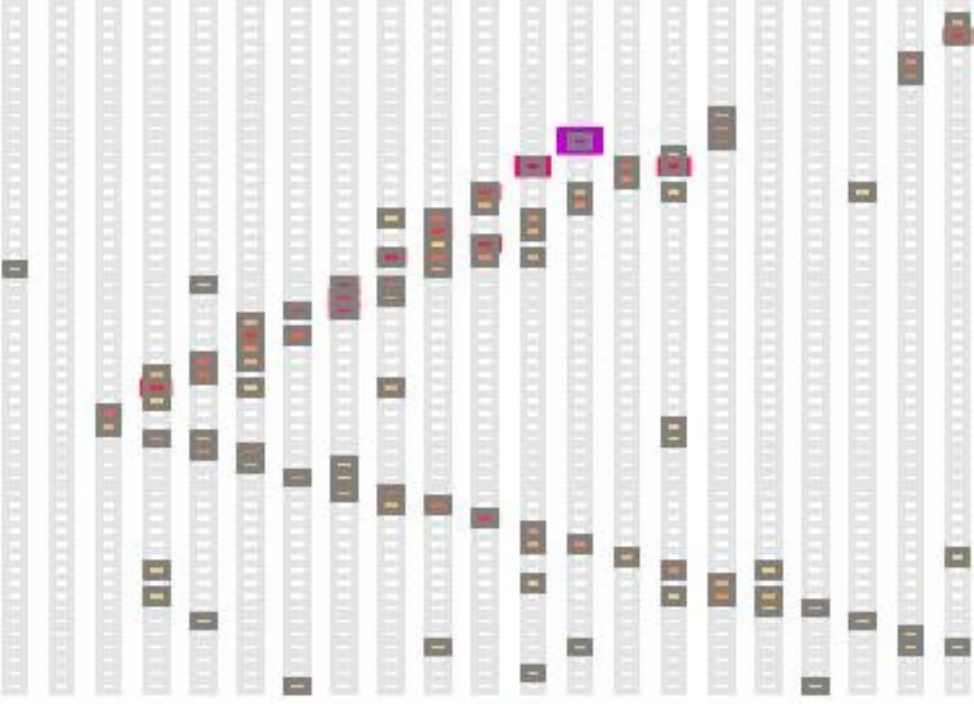}
    \caption{A CC event}
    \label{subfig:evdmrccCc}
  \end{subfigure}\qquad
  \begin{subfigure}[b]{0.27\textwidth}
    \includegraphics[width=\textwidth]{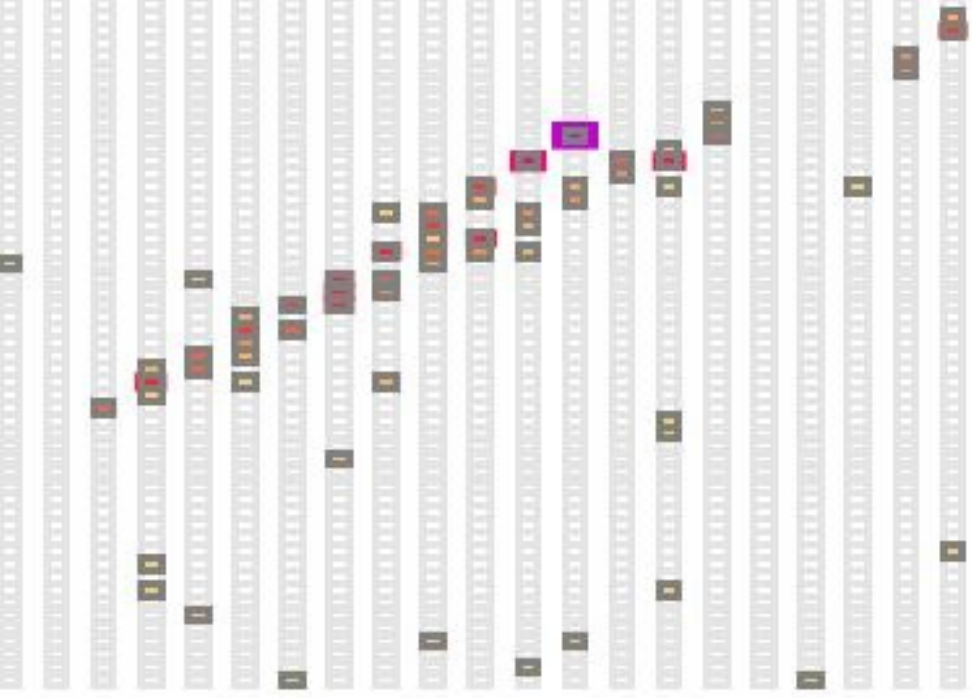}
    \caption{A MRCC event}
    \label{subfig:evdmrccMrcc}
  \end{subfigure}\qquad
  \begin{subfigure}[b]{0.27\textwidth}
    \includegraphics[width=\textwidth]{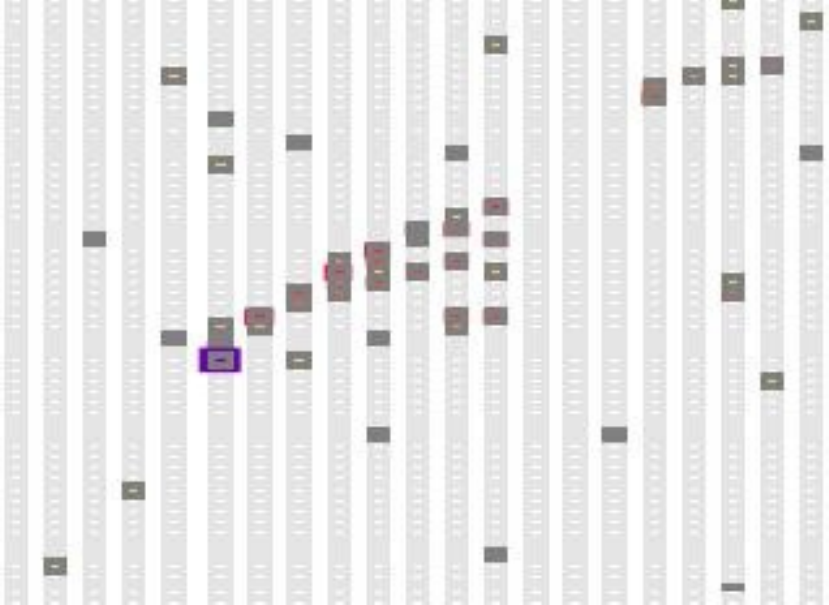}
    \caption{A NC event}
    \label{subfig:evdmrccNc}
  \end{subfigure}
  \caption{ Example topologies of the three types of events introduced in the text. \ref{subfig:evdmrccMrcc} is the muon removed version of the event in \ref{subfig:evdmrccCc}  }
  \label{fig:evdmrcc}
\end{figure}

Without the outgoing muon, a \numu CC interaction imitates a NC interaction since the outgoing neutrino in a NC interaction is invisible. This is only true if we can not resolve the differences in the hadronic showers resulting from the CC and the NC interactions. The removal of information about the outgoing muon in a \numu CC interaction produces what is known as a Muon-Removed Charged Current or MRCC event. Muon removal in data and Monte-Carlo provides an independent pseudo-NC sample that can be used to estimate the NC background in the \nue analysis. Figure \ref{fig:evdmrcc} shows a CC, a MRCC and a NC event.
\setcounter{subfigure}{0}

\subsection{Construction of MRCC Events\label{sec:mrcc}}
To remove the muon, the muon track is first identified using a muon particle identification (PID) algorithm. The muon PID in \nova is based on the log-likelihoods of dE/dx per plane that the track passes through. The hits that belong to the muon track are then removed from the event. However, close to the vertex of the neutrino interactions, there is significant hadronic activity due to the recoiling nucleus. It becomes likely then that some of the hits on the muon track have energy contribution from other particles. 
\begin{figure}[htb]
  \centering
  \includegraphics[width=0.5\textwidth]{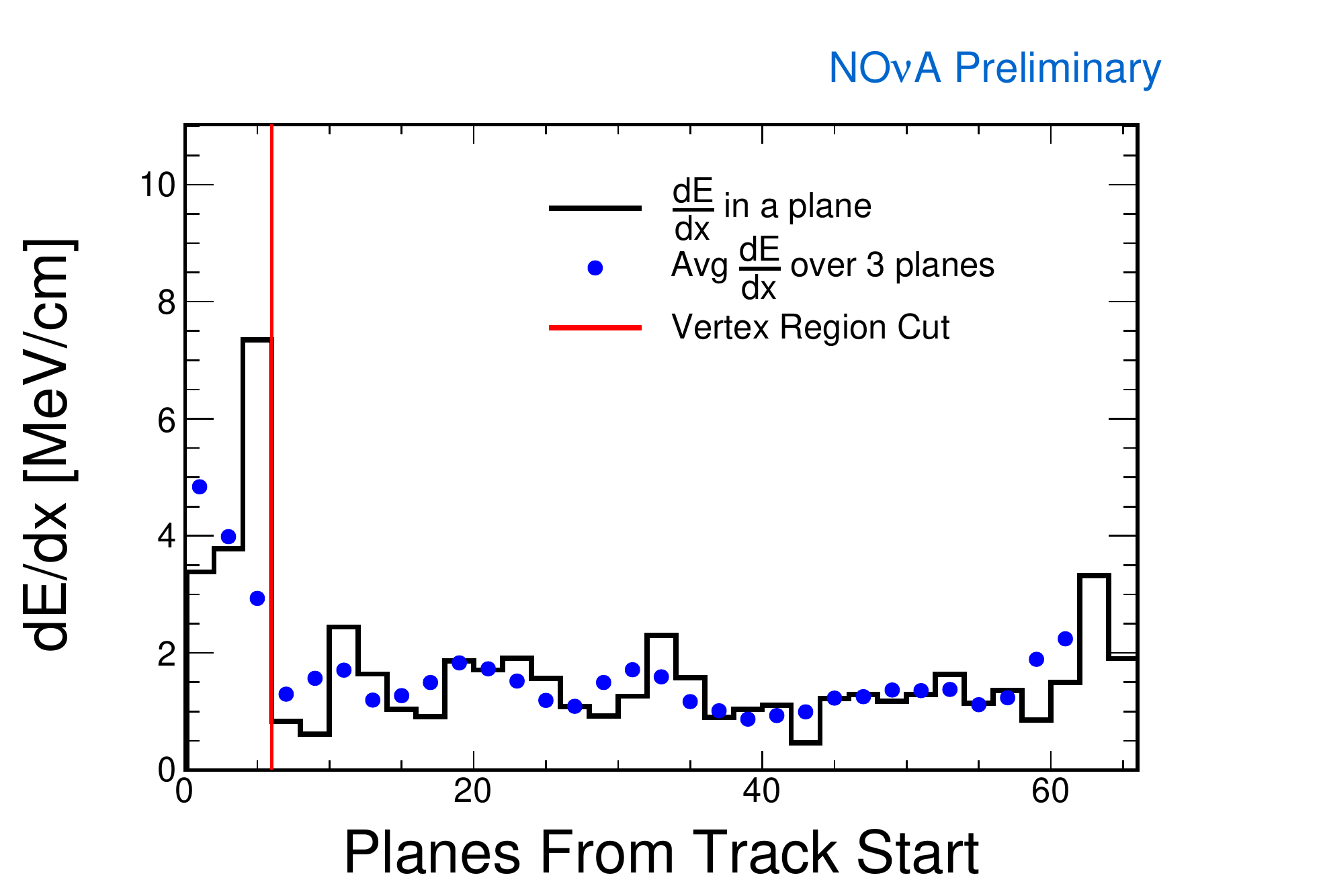}
  \caption{dE/dx profile of a muon in a \numu CC event. The red line marks the end of the vertex region to the left of which, the dE/dx is higher than that expected from a muon alone.}
  \label{fig:vert}
\end{figure}
 
To define the bound of the region of high hadronic activity, we use the dE/dx profile of the muon track. Muons are minimum ionizing particles (MIP) at energies typical in \nova ($\sim$1-4 GeV) \cite{Nakamura:2010zzi}
, and deposit $\sim 1.5$ MeV/cm in the \nova detectors. If a hadron coincides with a muon hit, the dE/dx in that plane is much higher than that of the muon alone. If the sliding average over three planes of dE/dx per plane on the muon track drops to a level consistent with a muon and stays there for the next three values of averaged dE/dx, then that plane marks the end of the region of hadronic activity. Figure \ref{fig:vert} shows an example of this technique. For hits within the vertex region, only one unit of MIP energy is removed, rather than removing the hit altogether. 

\subsection{Performance of Muon Removal}
\begin{figure}[h]
  \centering
  \begin{subfigure}[b]{0.47\textwidth}
    \includegraphics[width=\textwidth]{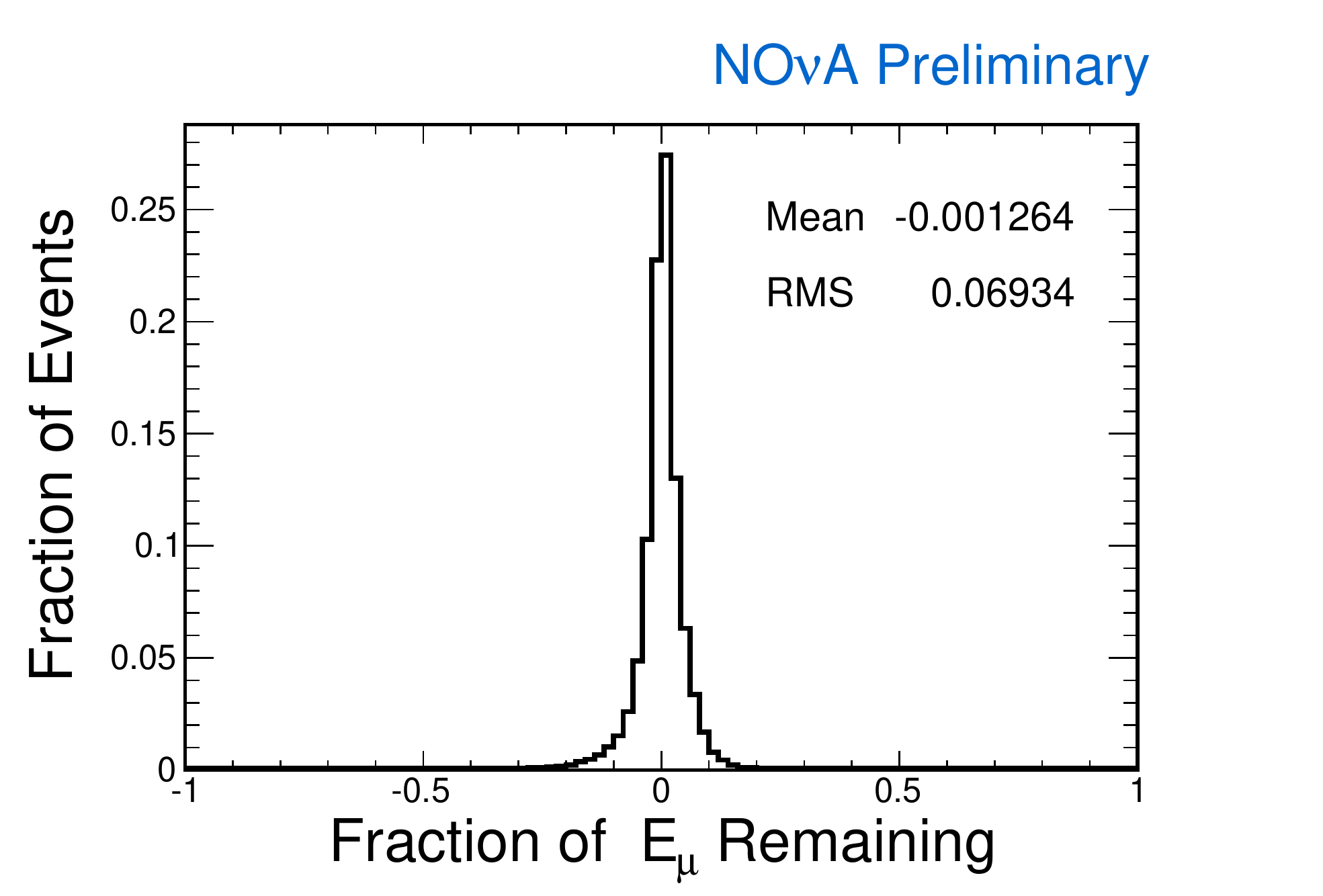}
    \caption{Fraction of muon left behind after removal, muFrac}
  \end{subfigure}\qquad
  \begin{subfigure}[b]{0.47\textwidth}
    \includegraphics[width=\textwidth]{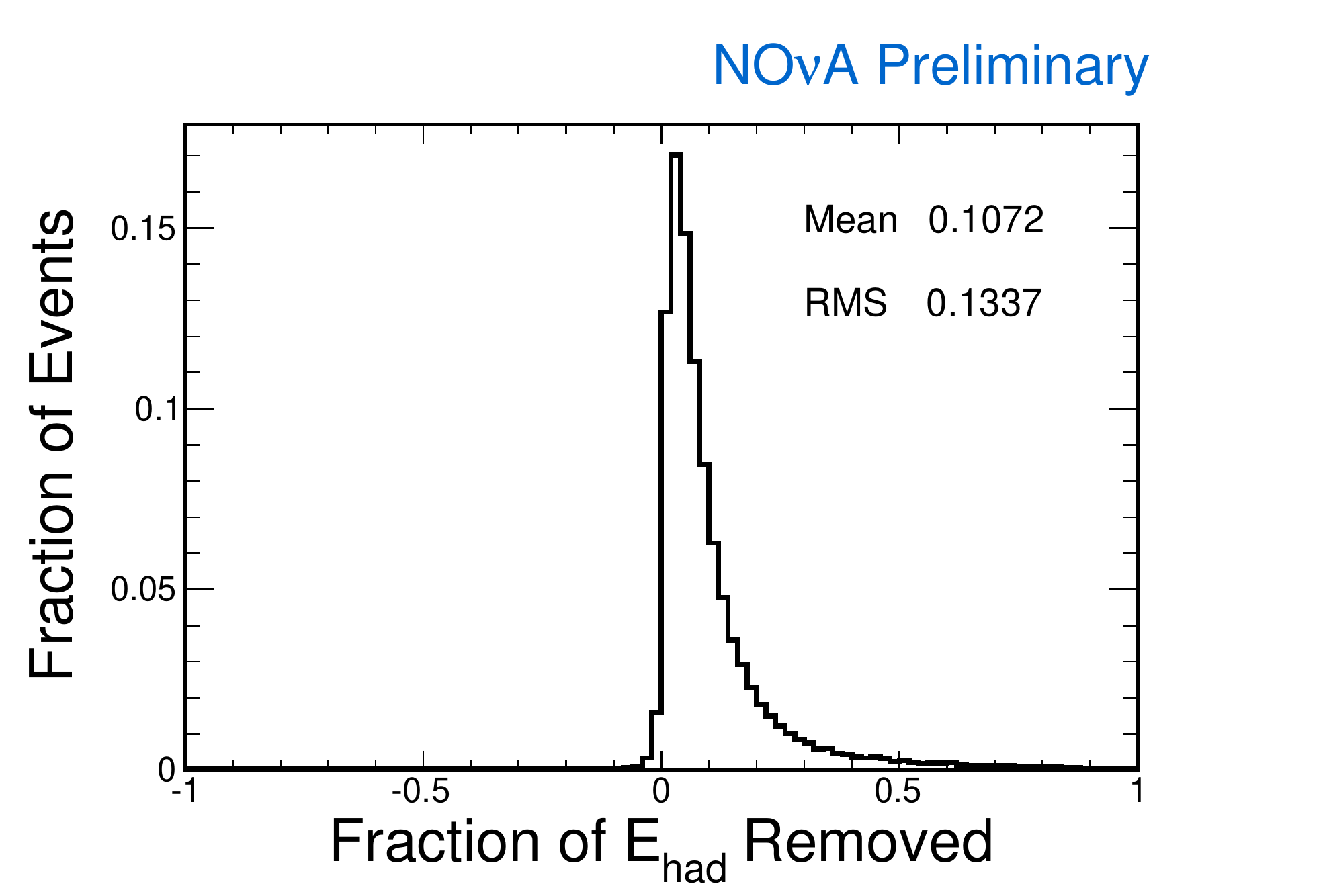}
    \caption{Fraction of hadronic energy removed in the process of muon removal}
  \end{subfigure}
  \caption{Variables to measure the performance of muon removal \label{fig:frac}}
\end{figure}

Since the purpose of muon removal is to remove the energy deposited by the muon in a \numu CC interaction and leave the hadronic shower energy untouched in the events. To test the process, two variables are defined. The first is the fraction of muon energy remaining in the event after muon removal and the second is the fraction of hadronic energy removed from the event during muon removal. If muon removal works perfectly, both these variables should be delta functions at 0. Figure \ref{fig:frac} shows the distributions of these variables. The tight peaks at $0$ indicates that muon removal is working well.

\setcounter{subfigure}{0}

\section{NC Background Estimation With MRCC}
\begin{figure}[h]
  \centering
  \begin{subfigure}[b]{0.47\textwidth}
    \includegraphics[width=\textwidth]{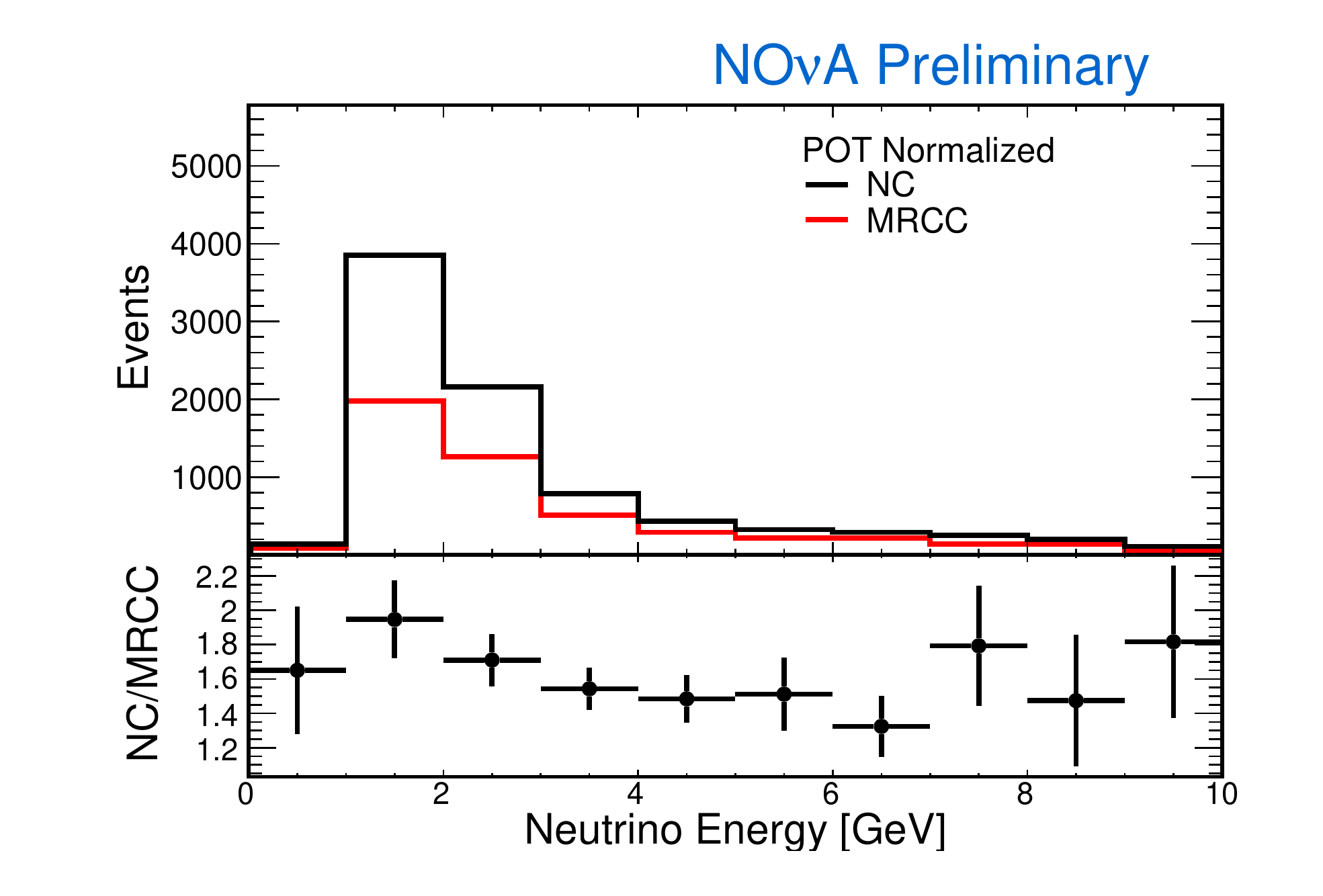}
  \end{subfigure}\quad
  \begin{subfigure}[b]{0.47\textwidth}
    \includegraphics[width=\textwidth]{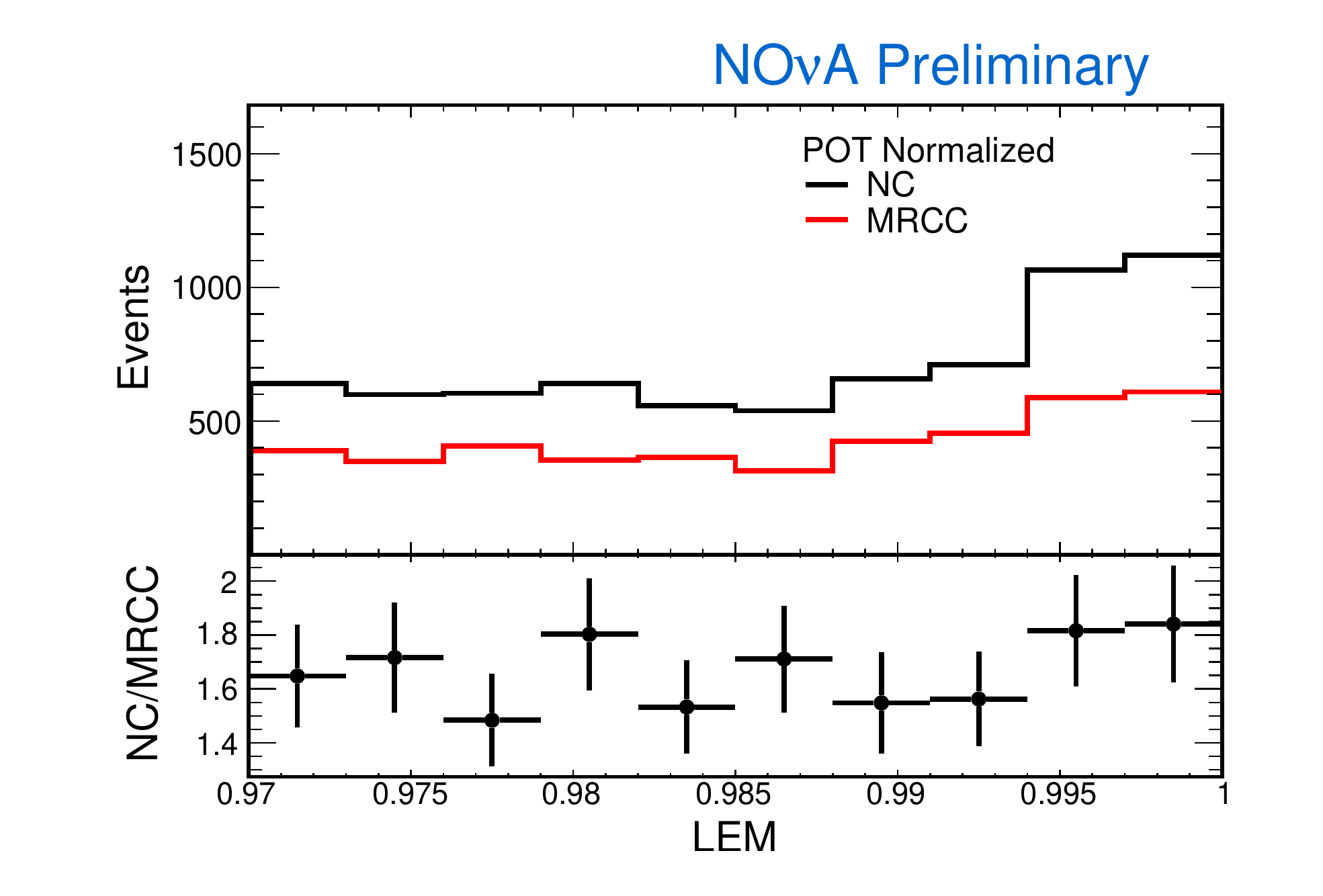}
  \end{subfigure}
\caption{Ratio NC/MRCC for various reconstructed quantities\label{fig:ratio} for events that pass all the \nue selection criteria in the near detector }
\end{figure}

MRCC provides a data-driven technique for studying hadronic showers that resemble NC events. MRCC can be performed on Monte-Carlo as well as data, and the two MRCC samples can then be used to predict the NC background. In this method, the muon candidate track from every neutrino interaction in Monte-Carlo and data is first removed as described in Section \ref{sec:mrcc}. These samples will henceforth be referred to as MRCC$_{MC}$ and MRCC$_{Data}$, respectively. The NC background event rate, NC$_{Pred}$ is predicted for each bin of a reconstructed variable like neutrino energy, PID value, longest track length etc, as given below:

\begin{equation}
  \label{eq:ncbg}
  \left( \text{NC}_{Pred} \right)_i = \left( \frac{ \text{NC}_{MC}}{ \text{MRCC}_{MC}}\right)_i \times \left( \text{MRCC}_{Data} \right)_i
\end{equation}
where $i$ refers to the bin index and $\text{NC}_{\text{MC}}$ is the true NC background selected as signal in the Monte-Carlo. Since many of the systematic effects that impact the NC and CC interactions in the same manner cancel in the ratio, this method results in a more precise estimate of NC background rate then a direct Monte-Carlo prediction. The distributions of the ratio variables in bins of reconstructed neutrino energy and Library Event Matching (LEM) \nue PID are shown in figure \ref{fig:ratio}.
\setcounter{subfigure}{0}

The hadronic showers resulting from CC interactions are however fundamentally different from NC showers. Such difference have been accounted for by considering the uncertainties on neutrino cross-section and interaction kinematics parameters that may impact CC and NC interactions differently. The systematic errors due to these effects have been incorporated in the error on the ratio in figure \ref{fig:ratio}. 

\begin{figure}
  \centering
  \begin{subfigure}[b]{0.49\textwidth}
    \vspace{-7pt}
    \includegraphics[width=1.15\textwidth]{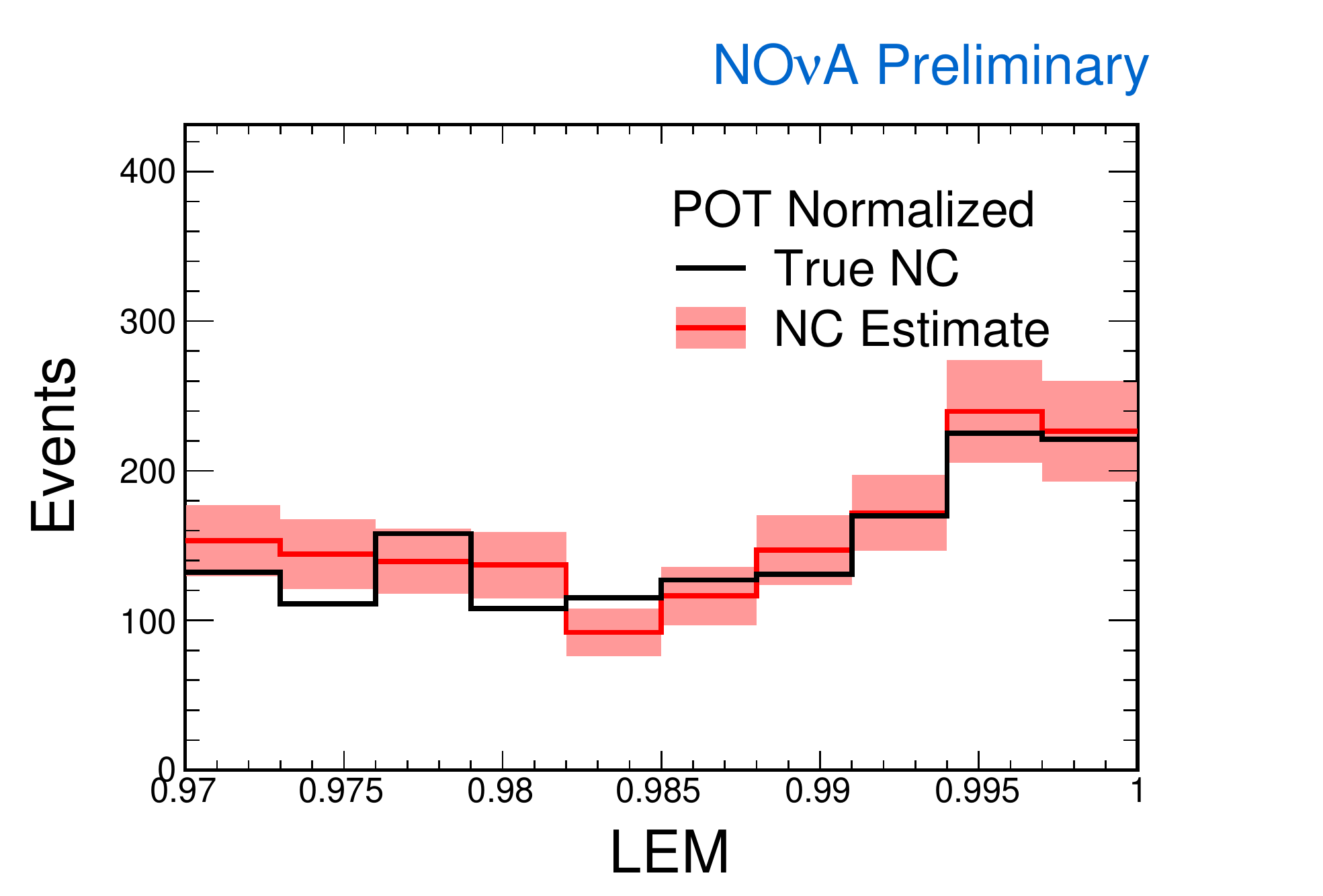}\hspace{-20pt}
  \end{subfigure}
  \begin{subfigure}[b]{0.49\textwidth}
    \vspace{-7pt}
    \includegraphics[width=1.15\textwidth]{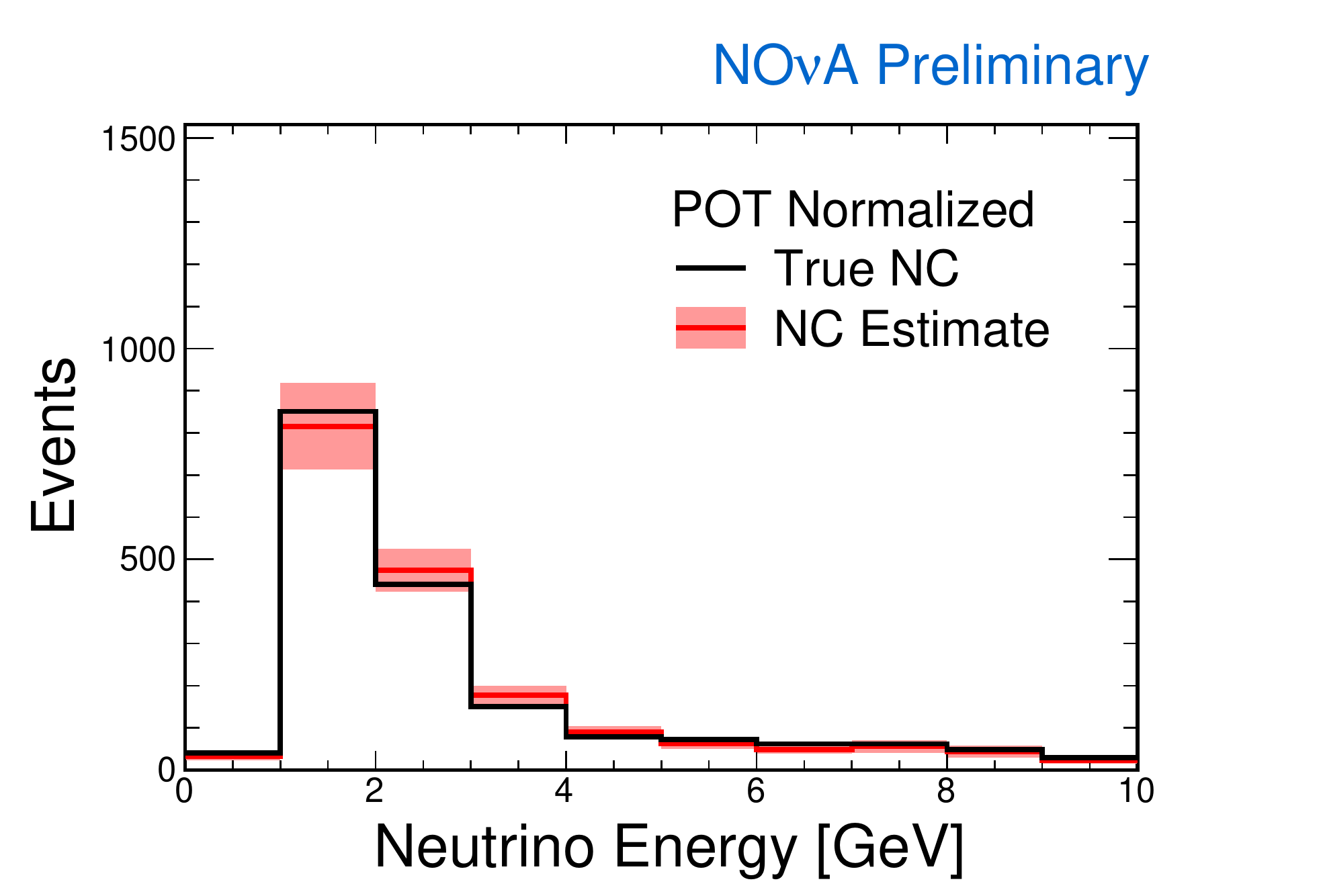}\hspace{-20pt}
  \end{subfigure}
  \caption{NC background estimate using MRCC method. The error band includes systematic and statistical errors \label{fig:est}}
\end{figure}

To test the performance of this method, we considered a statistically independent Monte-Carlo set as data and attempted to estimate the NC background rate in this set. The ratios from figure \ref{fig:ratio} were used to do this estimate and the results are presented in figure \ref{fig:est}. The MRCC method accurately reproduces the scale and shape of the true NC background.

\section{Conclusion}
A data-driven method to estimate NC background to the \nue oscillation analysis, using the near detector data, has been presented. The NC estimate from this method shows good agreement with the simulated rate in our tests with Monte Carlo.

\end{document}